\documentclass[twocolumn]{aastex631}

\usepackage[croatian,english]{babel}

\def \grb {\mbox{GRB\,230307A}}
\def \boat {\mbox{GRB\,221009A}}
\def \xmm {\emph{XMM-Newton}}

\def \cha {\emph{Chandra}}

\def \swi {\emph{Swift}}

\received{2023 July 25} \revised{2023 August 16} \accepted{2023 September 9}

\begin{document}

\title{\xmm\ and INTEGRAL observations of the bright \grb : vanishing of the local absorption and limits on the dust in the Magellanic Bridge}

\correspondingauthor{Sandro Mereghetti}
\email{sandro.mereghetti@inaf.it}

\author[0000-0003-3259-7801]{Sandro Mereghetti}
\affiliation{Istituto Nazionale di Astrofisica, Istituto di Astrofisica Spaziale e Fisica Cosmica di Milano, via A. Corti 12, 20133 Milano, Italy}

\author[0000-0001-6641-5450]{Michela Rigoselli}
\affiliation{Istituto Nazionale di Astrofisica, Istituto di Astrofisica Spaziale e Fisica Cosmica di Milano, via A. Corti 12, 20133 Milano, Italy}

\author[0000-0002-9393-8078]{Ruben Salvaterra}
\affiliation{Istituto Nazionale di Astrofisica, Istituto di Astrofisica Spaziale e Fisica Cosmica di Milano, via A. Corti 12, 20133 Milano, Italy}

\author[0000-0002-6038-1090]{Andrea Tiengo}
\affiliation{Scuola Universitaria Superiore IUSS Pavia, Piazza della Vittoria 15, 27100 Pavia, Italy}
\affiliation{Istituto Nazionale di Astrofisica, Istituto di Astrofisica Spaziale e Fisica Cosmica di Milano, via A. Corti 12, 20133 Milano, Italy}

\author[0009-0001-3911-9266]{Dominik P. Pacholski}
\affiliation{Istituto Nazionale di Astrofisica, Istituto di Astrofisica Spaziale e Fisica Cosmica di Milano, via A. Corti 12, 20133 Milano, Italy}
\affiliation{Dipartimento di Fisica G. Occhialini, Università degli Studi di Milano Bicocca, Piazza della Scienza 3, 20126 Milano, Italy}



\begin{abstract}

\grb\  is the second  brightest gamma ray burst detected in more than 50 years of observations and is located in the direction of the Magellanic Bridge. 
Despite its long duration,  it is most likely the result of the compact merger of a binary ejected from a galaxy in the local universe (redshift $z=0.065$).   
Our \xmm\   observation of its afterglow at 4.5 days shows a power-law spectrum with photon index $\Gamma =1.73 \pm0.10$, unabsorbed flux $F_{0.3-10\,\rm keV}=(8.8\pm0.5)\times 10^{-14}$ erg~cm$^{-2}$~s$^{-1}$ and no absorption in excess of that produced in our Galaxy and in the Magellanic Bridge. We derive a limit of $N_{\rm H}^{\rm HOST} < 5\times 10^{20}$ cm$^{-2}$ on the absorption at the GRB redshift, which is a factor $\sim\,$5 below the value measured during the prompt phase. 
We searched for the presence of dust scattering rings with negative results and set an upper limit of the order of $A_V<0.05$ on the absorption from dust in the Magellanic Bridge.
 
\end{abstract}

%\keywords{Gamma-ray bursts(629) --- Interstellar dust(836) --- Interstellar dust extinction(837) --- Interstellar scattering(854)}

\section{Introduction}\label{sec:intro}

Not many years after their discovery it was recognized that gamma-ray bursts (GRBs)  can be divided into two  groups, based mostly on their duration and, to a lesser extent, spectral hardness \citep{1984Natur.308..434N,1993ApJ...413L.101K}. It was later realized that this phenomenological distinction is related to different formation mechanisms.
GRBs of the ``long'' class have been firmly associated with type Ib/c core-collapse supernovae of massive stars 
\citep[e.g.][]{1998Natur.395..670G,2003Natur.423..847H,2003ApJ...591L..17S}.
The progenitors of the ``short'' GRBs remained elusive for a longer time, even though many hints consistently pointed to a compact binary merger progenitor 
\citep[e.g.][]{1992ApJ...395L..83N,1993Natur.361..236M,2007PhR...442..166N}. This was finally confirmed by the association of the short GRB~170817A with the gravitational wave signal GW~170817 produced by a binary neutron star merger \citep{Abbott2017a,Abbott2017b, 2017ApJ...848L..14G,2017ApJ...848L..15S}.

However, several results obtained in the last few years indicate that a classification based only on duration does not always reflect the intrinsic properties related to the GRB origin. 
For example, the short GRB 090426 resembled long bursts for what concerns its spectrum, energetics, and afterglow properties \citep{2009A&A...507L..45A,2011A&A...531L...6N}.
Evidence for a supernova, as usually detected in nearby collapsar-like events, was found for GRB 040924 and  GRB 200826A \citep{2008A&A...481..319W,Ahumada2021,Rossi2022}, which, although not extremely short, had rest-frame durations below 2 s.
On the other hand, deep observations of some long GRBs at low redshift failed to detect supernovae down to stringent limits \citep{2006Natur.444.1047F}, while kilonova signatures calling for a compact object merger origin have been associated with bursts of long duration, such as GRB 060614 \citep{Yang2015}  and, more recently, GRB 211221A \citep{Rastinejad2022,Troja2022,Yang2022}. A classification in terms of Type I/II, respectively for mergers and collapsars, seems thus more convenient \citep{2006Natur.444.1010Z}.

\begin{figure*}[ht!]
\centering
\includegraphics[angle=-90,trim=4cm 2cm 3cm 2cm,width=0.8\textwidth]{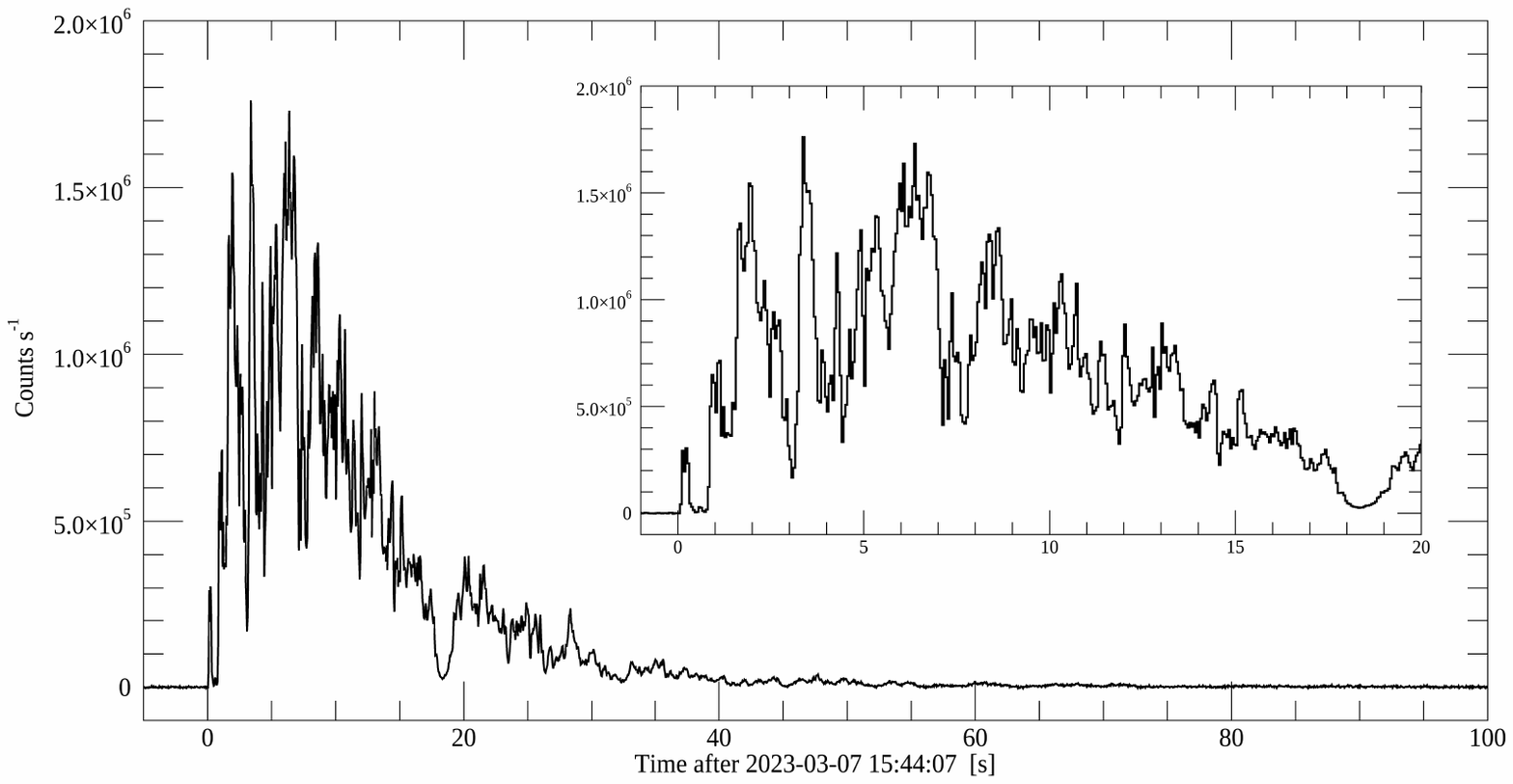}
\caption{Light curve of \grb\ at energies above 75 keV obtained with the SPI/ACS detector. The bin size is 50 ms. The inset shows a zoom of the initial part.} 
\label{fig:spiacs}
\end{figure*}

A recent example of a GRB with long duration  ($\sim$150 s), but most likely originating from a compact binary merger is provided by \grb\ \citep{2023arXiv230702098L,2023arXiv230702996D,2023arXiv230705689S,2023arXiv230311083W}. 
The evidence for a Type I classification  comes from  the James Webb Space Telescope discovery of an associated  kilonova \citep{2023arXiv230702098L,2023arXiv230800638Y}, similar to that seen for the gravitational wave merger GW~170817 \citep{Abbott2017a}. Most likely, the binary merger was ejected from a galaxy seen at an angular offset of 30$''$ from \grb . Given the galaxy's redshift of $z=0.065$, this corresponds to a large, but still plausible, projected distance of about  40 kpc.
The alternative association  with a galaxy  at redshift $z=3.87$ \citep{2023GCN.33580....1L}  is unlikely because it would imply an unprecedented isotropic energy as large as  $10^{56}$ erg  and,  at such a high redshift,  the observed rapidly varying kilonova-like emission  corresponds to   rest-frame rise and decay times too short for a supernova of this luminosity.  
Furthermore, the spectroscopic detection of an emission line at 2.15 $\mu$m,     interpreted as due to tellurium, makes the low redshift  type I GRB hypothesis very strong \citep{2023arXiv230702098L}.

\grb\ immediately attracted much interest owing to its very large fluence of a few $10^{-3}$ erg cm$^{-2}$
\citep{2023GCN.33407....1D,2023GCN.33411....1D}, surpassed only by that of the exceptional GRB 221009A\footnote{Although the occurrence within six months of the two brightest GRBs ever observed 
seems at first remarkable, the time distribution of the complete sample of 17 GRBs with fluence above $10^{-3}$ erg cm$^{-2}$ detected since  1971 (Table 2 of \citealt{2023ApJ...946L..31B}) is fully consistent  with a constant rate.}  \citep{2023GCN.33406....1X,2023GCN.33414....1B,2023GCN.33427....1S}. 
\grb\ is also interesting because it occurred in the direction of the Magellanic Bridge, a structure  consisting mainly of HI which connects the Small and Large Magellanic Clouds and was probably produced by the tidal interaction between these two galaxies  \citep[see, e.g.,][and references therein]{2021A&A...649A...7G}. The occurrence of a bright GRB in this direction offers the possibility to constrain the poorly known properties of the interstellar dust in the Bridge  through the analysis of X-ray scattered radiation, as done  for Galactic dust clouds with other GRBs     \citep{vaughan04,2006A&A...449..203T,2006ApJ...639..323V,2007A&A...473..423V,2017MNRAS.472.1465P,2023ApJ...946L..30T}.

Here we report the results of an \xmm\   observation  of \grb\ carried out about 5 days after the burst occurrence (Section~\ref{sec:xmm}), complemented by data of the prompt emission in the hard X-ray range obtained with the INTEGRAL satellite (Section~\ref{sec:integral}).  In Section~\ref{sec:dust} we use the  X-ray data to look for the presence of dust scattering rings and to derive constraints on the amount of dust in the Magellanic Bridge.

\section{Data analysis and results} 

\subsection{Prompt gamma-ray emission with INTEGRAL} \label{sec:integral}

\grb\ was detected by the Anti-Coincidence Shield (ACS) of the SPI instrument on board the INTEGRAL satellite.  
The ACS, besides serving as an active shield for the germanium  detectors of SPI, acts as a nearly omni-directional detector with high sensitivity at energies above $\sim\,$75 keV \citep{2003A&A...411L.299V}. It provides data with fixed time resolution of 50 ms in a single energy channel and without directional information. The burst occurred at an angle of 75$^{\circ}$ with respect to the INTEGRAL pointing direction, and at an azimuthal angle for which the ACS is not obstructed by other instruments on board the satellite. Therefore, the ACS provided an optimal response for the direction of \grb\  \citep[see, e.g.,][]{2017A&A...603A..46S}.

The ACS light curve of \grb\ is presented in Figure~\ref{fig:spiacs}. The burst showed remarkable variability on short time scales, within an envelope with a fast-rise  and a slower decay. The $T_{90}$  duration is of 28.6 s, but significant emission is visible up to 130 s after   $T_0=$ 15:44:07 UT.  
The burst starts with a $\sim\,$0.2 s long precursor, which contains $\sim\,$0.4\% of the total fluence. This is followed by a multipeaked pulse lasting about 17 s, until a dip at $\sim T_0+18$ s.
The temporal decay after $T_0+20$ s can be approximated by a power law with index $\alpha=-3.467\pm0.007$.
The total fluence, from $T_0$ to $T_0+130$ s, is $(1.5239\pm0.0008)\times10^7$ ACS counts, which, adopting the average conversion factor of 1 ACS count $\sim10^{-10}$ erg cm$^{-2}$ \citep{2009essu.confE..49V}, corresponds to $\sim1.52\times10^{-3}$ erg cm$^{-2}$ in the 75--1000 keV  range.

\subsection{X-ray afterglow with \xmm}
\label{sec:xmm}

A 50 ks long Target of Opportunity observation of \grb\ was carried out with the \xmm\ satellite, starting on 2023 March 12 at 02:16:54 UT, about 4.5 days after the GRB trigger. 
The EPIC-pn \citep{struder01} and the two EPIC-MOS \citep{turner01} cameras were operated in Full window mode, with the thin optical-blocking filter. 
After processing the data with the  \texttt{SAS 19.1.3} \citep{gabriel04} and the most recent calibration files, we selected the EPIC events with standard filtering expressions  and  removing time intervals of high background. This resulted in net exposure times of 37.2 ks (pn) and 41.5 ks (MOS). In addition to the GRB afterglow, several sources were detected (see Appendix~\ref{images}).

The  afterglow spectra were extracted from a circular region of 30$''$ radius, centered at coordinates R.A.$=04^{\rm h} 03^{\rm m} 26^{\rm s}\!\!.24$, Dec.$=-75^\circ 22' 43''\!\!\!.\,8$, while those of the background from a nearby circle of 40$''$ radius. The spectra  of the three cameras were  rebinned with a minimum of 30 counts per bin and then fitted simultaneously  using the \textsc{xspec} software (version 12.13.0c). For the  absorption model we used cross sections and abundances from \citet{2000ApJ...542..914W}.  Errors on the spectral parameters are given at 1$\sigma$ confidence level. 

\begin{figure}
    \centering
\includegraphics[width=0.5\textwidth]{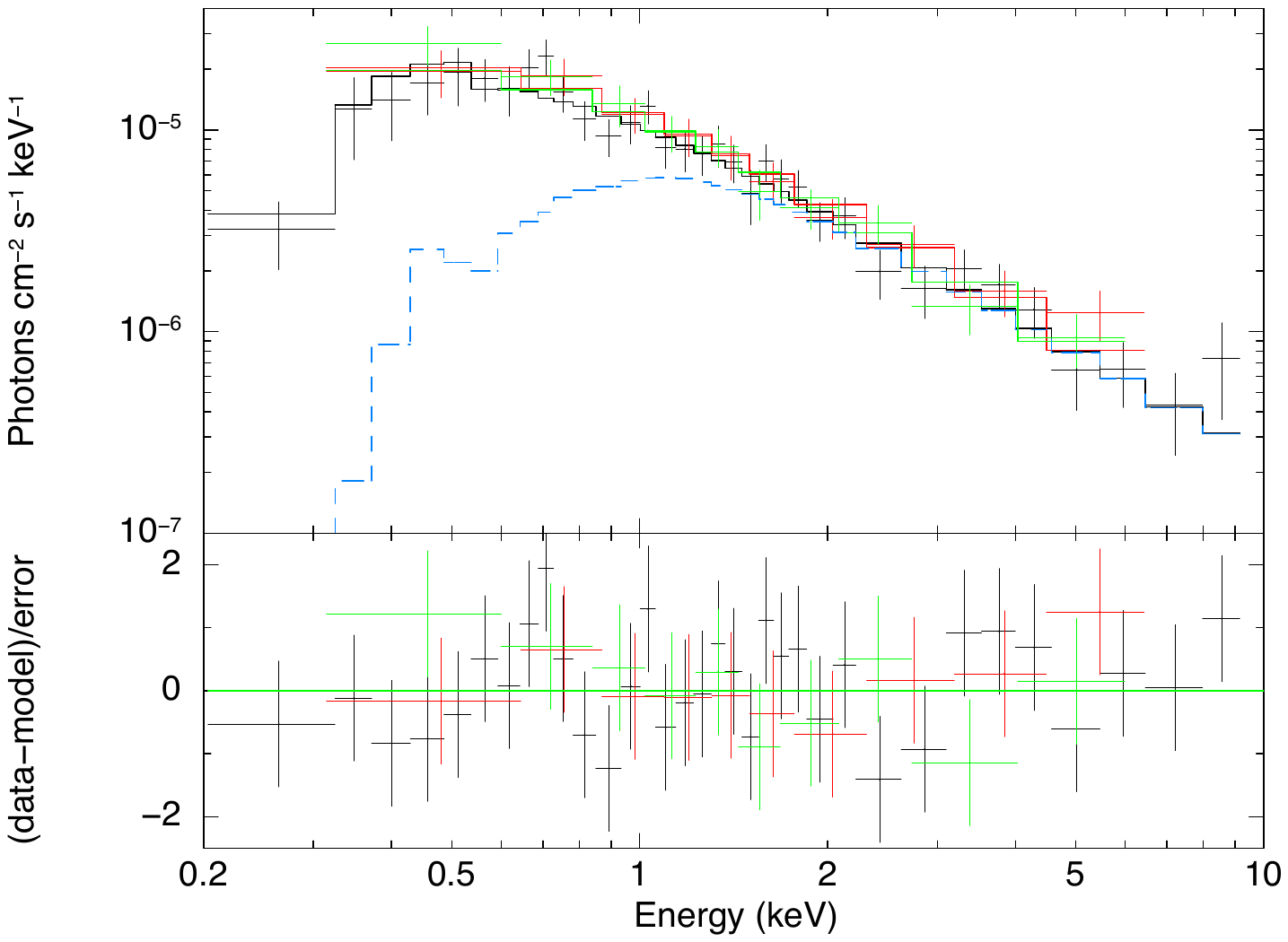}
    \caption{Top panel: EPIC-pn (black), -MOS1 (red) and -MOS2 (green) spectra of the afterglow of \grb\ fitted with an absorbed power law.  The blue dashed line indicates, for comparison, the spectrum expected with the intrinsic absorption $N_{\rm H}^{\rm HOST}=2.73\times10^{21}$~cm$^{-2}$   measured in the prompt emission. Bottom panel:  residuals of the best fit in units of $\sigma$.}
    \label{fig:spec}
\end{figure}

The spectra are well described ($\chi^2/dof = 29.14/49$, null hypothesis probability (nhp) = 0.99) by an absorbed power law with photon index $\Gamma = 1.73 \pm 0.10$, column density $N_{\rm H} = (9 \pm 2)\times 10^{20}$~cm$^{-2}$, and unabsorbed flux   $F_{0.3-10\,\rm keV}=(8.8\pm0.5)\times 10^{-14}$~erg~cm$^{-2}$~s$^{-1}$ (see Figure~\ref{fig:spec}).

\begin{figure}
    \centering
    \includegraphics[width=0.5\textwidth]{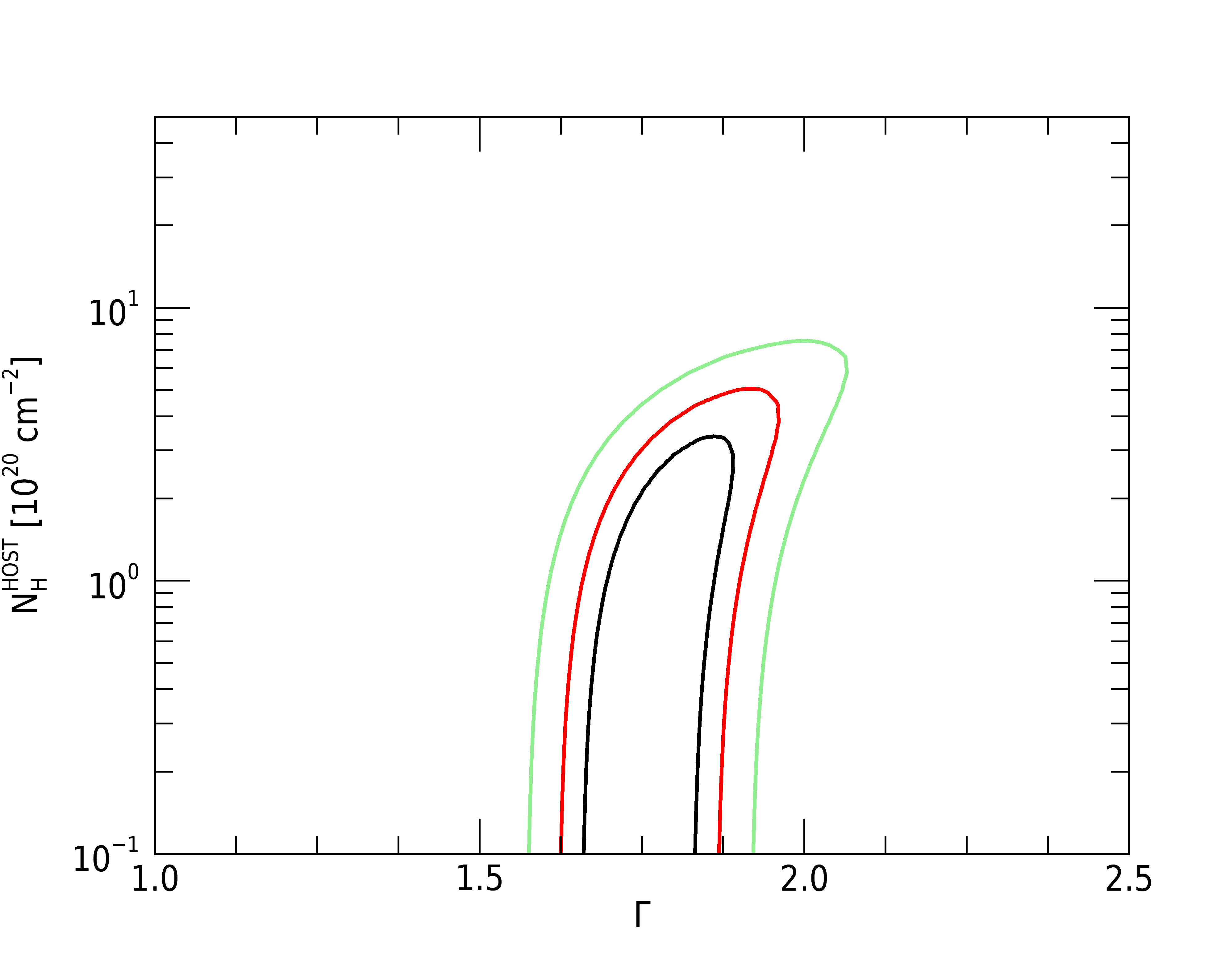}
    \caption{Confidence contours (1, 2 and  3$\sigma$) of photon index $\Gamma$ and absorption at the GRB redshift $N_{\rm H}^{\rm HOST}$.}
    \label{fig:cont}
\end{figure}

The derived absorption is consistent with the local (Galactic plus Magellanic Clouds) column density in this direction,  $N_{\rm H}^{\rm LOC}=9.4\times10^{20}$~cm$^{-2}$  \citep{2016A&A...594A.116H}.
Therefore, in order  to constrain   possible  absorption in the GRB host,  we fixed the column density to $N_{\rm H}^{\rm LOC}$ and added to the model a redshifted absorption component ($N_{\rm H}^{\rm HOST}$).
We found that, at the GRB redshift  $z=0.065$, the host absorption is $N_{\rm H}^{\rm HOST}<5\times10^{20}$~cm$^{-2}$ 
($2\sigma$ upper limit, see Figure \ref{fig:cont}).

\begin{figure}
    \centering
    \includegraphics[width=0.5\textwidth]{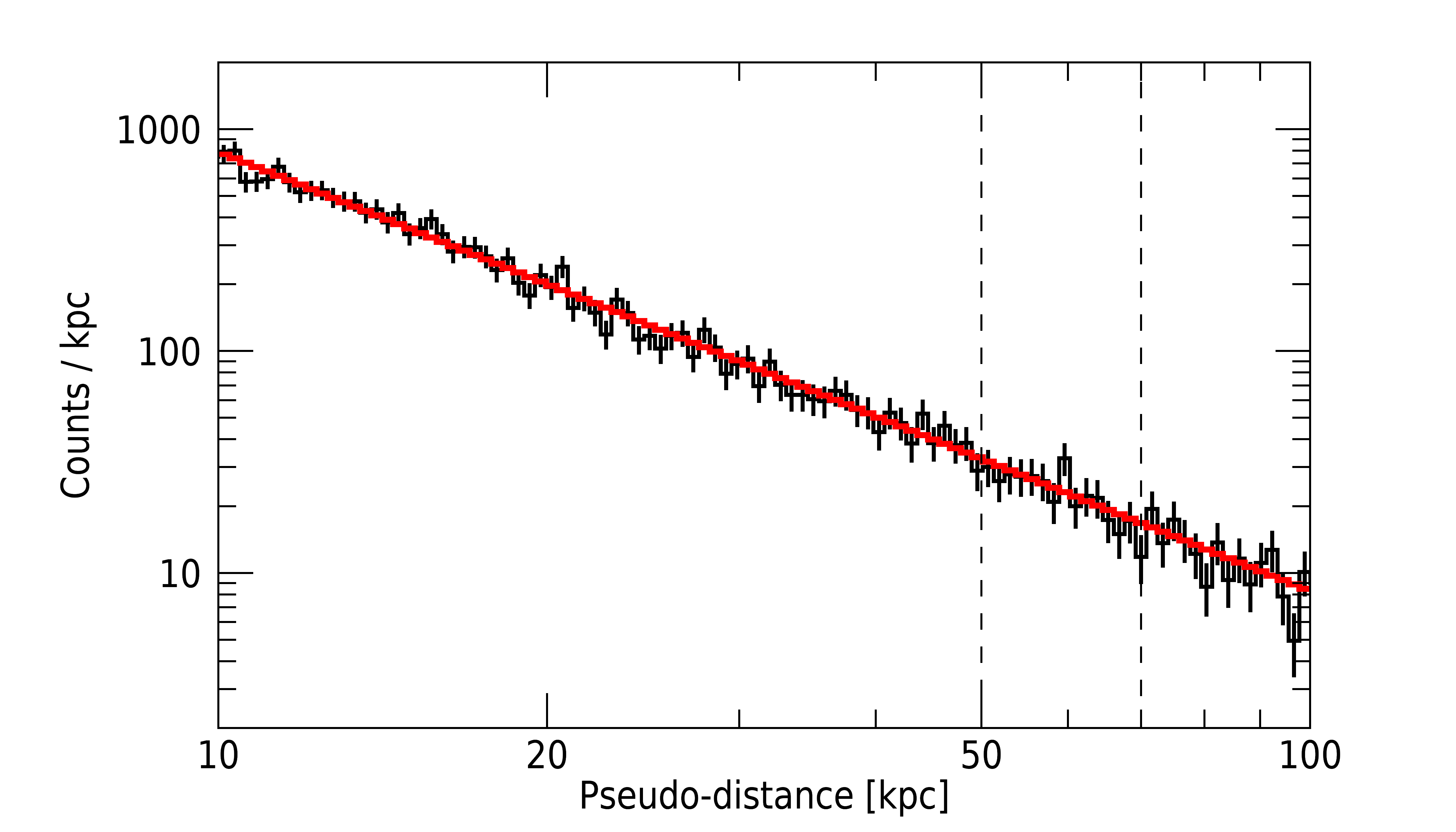}
    \caption{Pseudo-distance distribution extracted from EPIC-pn and -MOS events in the 0.5--4 keV energy range. The red line represents the best power-law fit to the data, while the vertical dashed lines indicate the relevant distance range for the analysis of  interstellar dust in the Magellanic Bridge.}
    \label{fig:pseudoD}
\end{figure}

\begin{figure*}
    \centering
    \includegraphics[width=1\textwidth]{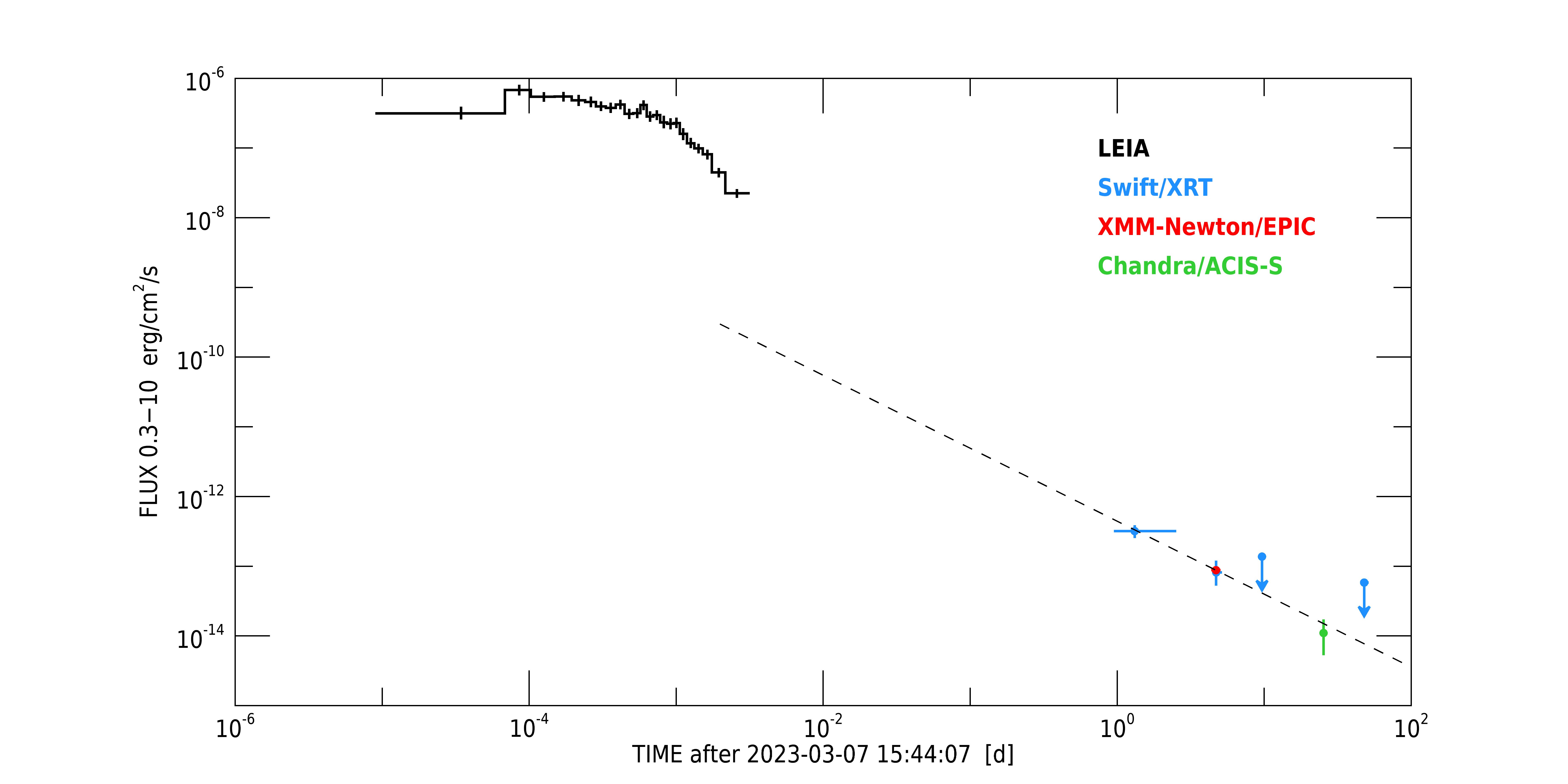}
    \caption{X-ray flux between 0.3--10 keV of \grb\ (black histogram: LEIA) and its afterglow (blue: \swi/XRT; red: \xmm/EPIC; green: \cha/ACIS-S). The light curve of the afterglow has been fitted with      a power law in the range $T_0 +[1.3,25.3]$ d (dashed line).}
    \label{fig:afterglow}
\end{figure*}

\subsection{Limits on dust in the Magellanic Bridge}
\label{sec:dust}

We searched for the presence of dust scattered events using the method based on ``pseudo-distances''  \citep{2006A&A...449..203T}, which provides a higher sensitivity for expanding rings compared to a simple analysis of time-integrated images.
Briefly, this method exploits the fact that for a source at distance much greater than that of the dust, X-ray photons detected at an angle $\Theta$ from the source direction arrive with a time delay $t=(D/2c)\Theta^2$,   where $D$ is the distance of the dust. A pseudo-distance $D_i$ can be computed  for each detected count,  based on its coordinates ($x_i$, $y_i$) and arrival time $t_i$:

\begin{equation}
D_i =\frac{ 2c\, (t_i-t_B)}{(x_i-x_B)^2+(y_i-y_B)^2}
\end{equation}

\noindent
where $x_B$, $y_B$ and $t_B$ are the coordinates and time of the GRB. An expanding X-ray ring produced by dust at distance $D$ appears as a Lorentzian peak in the $D_i$ distribution.

To derive the $D_i$ distribution we selected the \xmm/EPIC events with energy in the range 0.5--4 keV, \textsc{flag}$\,=\,$0, \textsc{pattern}$\,\leq\,$4 (pn) and $\leq\,$12 (MOS), and we removed several point sources (see Appendix~\ref{images}) by excluding circular regions of
radius $\sim$\,20$''$. The resulting empty regions, as well as the dead areas between CCD gaps, were filled with uniformly-distributed events, in order to avoid artifacts and spurious peaks  in the $D_i$ distribution. 

Most of the Galactic dust in this direction is concentrated in two layers at distances of 225 pc and 375 pc \citep{lallement22}, but, at the time of our   observation, possible scattering rings produced by these layers had already expanded to angular radii  of $\sim19'$ and   $\sim14.5'$, outside the EPIC field of view.   On the other hand, X-ray scattered by dust in the Magellanic Bridge can be detected because radii in the range $\sim 1-2'$ are expected in this case  (see Figure \ref{fig:image}).

As it can be seen in Figure~\ref{fig:pseudoD}, the distribution of the $D_i$ values does not show any prominent peak  in the range of relevant distances.
In fact it is well fit by a single power law with slope $-1.98\pm0.02$  ($\chi^2$/dof $=1.05$,   nhp = 0.35), as expected for uniformly distributed background counts and unscattered X-ray photons.

\begin{figure}[ht!]
    \centering
    \includegraphics[width=0.5\textwidth]{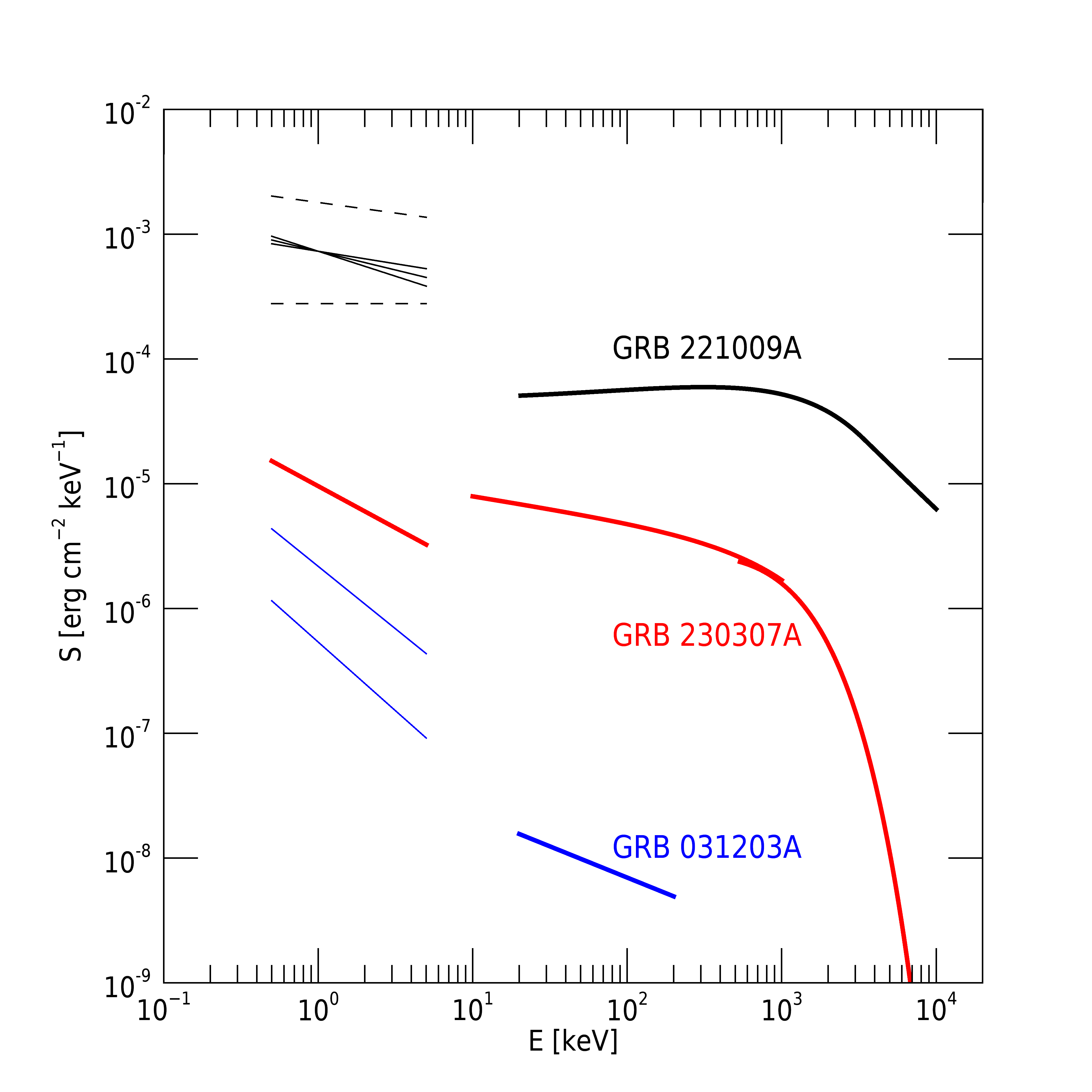}
    \caption{Broad band fluence spectra for three GRBs with detections or limits on the dust scattering. 
    The spectra for \grb\ (red) are from \citet{2023arXiv230705689S}.
    For \boat\ (black) we indicate the best estimate (solid) and the maximum and minimum values (dashed) for the soft X-ray fluence derived from the analysis of dust scattering rings \citep{2023ApJ...946L..30T}, while the gamma-ray spectrum is from \citet{2023ApJ...949L...7F}. 
    For GRB 031203 (blue) we indicate two estimates of the soft X-ray fluence from dust scattering analysis by  \citet{2006A&A...449..203T}  (lowest) and \citet{2006ApJ...636..967W} (highest), while the gamma-ray spectrum is from \citet{2009A&A...495.1005V}.}
    \label{fig:2grb}
\end{figure}

To derive upper limits on the scattered X-ray flux we carried out  Monte Carlo simulations in which we added to the observed $D_i$ histogram $N_s$ photons distributed according to Lorentzian profiles expected for different distances  of the dust.
In principle, the Lorentzian width $\Gamma$ depends on the thickness $\delta D$ of the dust layer, but, in the case at hand, it is dominated by the finite angular resolution of the instrument ($\delta \Theta_{\rm FWHM} \sim 5''$), which results in $\Gamma = 2 D\, | \delta \Theta_{\rm FWHM}/\Theta |$. At the mean time of the observation (4.7 d after the burst) this gives  the value $\Gamma \sim 0.017  D_{\rm kpc}^{3/2}$ kpc, which we  used in our simulations. 
These were carried out for two representative values, 50 and 70 kpc, which bracket possible distances for dust structures in the Magellanic Bridge \citep[see, e.g.][]{2016AcA....66..149J}.
We found 95\% confidence level upper limits of
$N_s = 245$ EPIC counts for $D=50$ kpc ($\Gamma=6$ kpc) and  $N_s = 260$ EPIC counts for $D=70$ kpc ($\Gamma=10$ kpc).

\section{Discussion}

The X-ray flux (0.3--10 keV)  of our \xmm\ observation is plotted in Figure~\ref{fig:afterglow}, together with  those from other available measurements of the X-ray afterglow. These are the \swi/XRT observations carried out at $T_0+1.3$~d \citep{2023GCN.33465....1B}\footnote{See also https://www.swift.ac.uk/xrt\_products/TILED\_GRB00110/ Source2/curve and https://www.swift.ac.uk/xrt\_curves/00021537/}   and $T_0+4.7$~d, and the \cha\ observation at $T_0+25.3$~d \citep{2023GCN.33558....1R}. 
We also show in Figure~\ref{fig:afterglow} the  X-ray light curve of the prompt emission   measured with LEIA in the 0.5--4 keV \citep{2023arXiv230705689S}, opportunely rescaled in the  0.3--10 keV range according to their spectrum (see below).
The temporal evolution  of the afterglow is well fitted by a power law of index $\alpha=-1.05\pm0.13$ and normalization $n = (4.4\pm0.9)\times10^{-13}$~erg~cm$^{-2}$~s$^{-1}$ at $T_0+1$ d.

The  upper limit we derived at $T_0+4.5$ d on the host absorption, $N_{\rm H}^{\rm HOST}<5\times10^{20}$~cm$^{-2}$, is significantly smaller than the value observed during the prompt emission phase, $N_{\rm H}^{\rm PROMPT}=2.73\times10^{21}$~cm$^{-2}$ \citep{2023arXiv230705689S}. 
Evidence of X-ray column density variations has been reported so far only for a handful of GRBs \citep{Starling2005,Campana2007,Campana2021,Grupe2007}. Indeed, a systematic analysis of 199 GRBs 
showed that only seven bursts display signs of a decrease in the intrinsic $N_{\rm H}$ \citep{Valan2023}, although the observed variation could also be explained in terms of spectral evolution of the afterglow emission.
Variations in the X-ray absorption are  predicted due to the ionization of the surrounding medium by the GRB radiation \citep{Perna1998,Lazzati2002,Perna2002}.  However, it is unlikely that this scenario applies to the case at hand: at more than 40 kpc from its host galaxy, \grb\ probably exploded in a low density environment, as also suggested by the relative faintness of its afterglow \citep{2023arXiv230800638Y,2023arXiv230702098L}. 
The absorption measured with LEIA during the prompt phase was most likely caused by very local material, possibly ejected during the last phases preceding the merger. The ejecta responsible for the absorption might have moved out of the line of sight by the time of the \xmm\ observation. 
Another possibility is that the afterglow was produced at a large distance, not affected by    material close to the central engine. 
A fully self-consistent model of mass ejection in short GRBs   would  be a significant step forward, but unfortunately this has not been developed yet.
In this respect, more measurements of the X-ray absorption and its evolution during the initial phases of the burst would be very useful.

We searched for X-ray scattering rings caused by dust in the Magellanic Bridge, with negative results.  Assuming a model for the scattering cross section and knowing the X-ray fluence and spectrum of \grb , the upper limits on the number of scattered photons derived in Section \ref{sec:dust} can be translated into limits on the amount of dust. 
We adopt the analytical dust model of \citet{draine03} and the results obtained with the LEIA experiment \citep{2023arXiv230705689S} on the X-ray prompt emission (a power-law spectrum with photon index $\Gamma=1.67$, 
$N_{\rm H}^{\rm LOC}=9.41\times10^{20}$~cm$^{-2}$,  $N_{\rm H}^{\rm HOST}=2.73\times10^{21}$~cm$^{-2}$ and fluence $F_{\rm 0.5-4\,keV}$ = $2.27\times10^{-5}$ erg cm$^{-2}$).  
The resulting limits of $A_V<0.040-0.056$, for dust at distances in the 50--70 kpc range,  show the potentiality of this method to constrain the dust properties in the Magellanic Bridge. 
However, we note that the results depend on the chosen dust model. As an example, adopting the  BARE-GR-B model of \citet{zubko04},  which provided the best fit in other observations of dust scattering X-ray halos/rings (e.g., \citealt{rsmith06,tiengo10,2023ApJ...946L..30T}), we can limit the quantity of dust in this part of the Magellanic Bridge to $A_V<0.25-0.32$.
Even so, these limits are more constraining than the total reddening $E(B-V)=0.08\pm0.08$ reported in the \citet{2022MNRAS.511.1317C} map, which in this direction has a spatial resolution of 55$^{\prime}$ and estimates the same reddening value also for foreground Galactic stars.

\section{Conclusions}

We have observed  \grb\ with \xmm\  in order to study its X-ray afterglow about five days after the burst and to search for dust scattering features, exploiting its very high brightness and fortuitous location close to the Large Magellanic Cloud. 

The good counting statistics and wide energy range of the EPIC spectra allowed us to reveal a significant decrease in the absorption, which passed from $2.73\times10^{21}$~cm$^{-2}$  in the prompt emission phase to less than $5\times10^{20}$~cm$^{-2}$ during the afterglow at $\sim$4.5 days after the burst.   
This large variation suggests that tne absorption in the prompt emission was caused by ejecta very close to the central engine, which moved out of the line of sight by the time of the \xmm\ observation.

The study of dust scattered X-rays from GRBs with sensitive telescopes offers great potentialities.  As it is schematically shown in Figure~\ref{fig:2grb}, in at least two cases, GRB 031203A \citep{2006ApJ...636..967W,2006A&A...449..203T} and \boat\ \citep{2023ApJ...946L..30T},   these analysis provided evidence for a bright soft X-ray excess that could not be observed directly.
Conversely, the prompt soft X-ray emission of \grb\ was directly observed and we could  exploit the resulting information to set limits on the amount of dust in the  Magellanic Bridge. 
Although the limits are not very deep in this particular case, our results indicate that more constraining observations are within reach of the current instrumentation. For example, an earlier  \xmm\ observation of \grb ,  performed when the X-rays potentially scattered by nearby dust clouds were still within the telescope field of view, would have allowed to disentangle the effects of dust in our Galaxy and in the  Magellanic Bridge.

%% The "ht!" tells LaTeX to put the figure "here" first, at the "top" next
%% and to override the normal way of calculating a float position
%\begin{figure}[ht!]
%%\plotone{samplefig.png}
%\caption{The cost for an author to publish an article has trended downward over time. This figure shows the average cost of an article from 1990 to 2020 in 2021 adjusted dollars. 
%\label{fig:general}}
%\end{figure}

%Figure \ref{fig:general} shows the changes in the author publication charges (APCs) from 1990 to 2020 in the AAS Journals.  

%% Also note that the akcnowlodgment environment does not support long amounts of text. If you have a lot of people and institutions to acknowledge, do not use this command. Instead, create a new 

\section{Acknowledgments}
The scientific results reported in this article are based on observations obtained with \xmm\ and INTEGRAL, ESA science missions with  instruments and contributions directly funded by ESA Member States,  NASA and Russia.
We acknowledge financial support from the Italian Ministry for University and Research, through grants 2017LJ39LM (UNIAM) and INAF Large Program   for Fundamental Research 2022. 
We thank the \xmm\ Mission Scientist N. Schartel for granting the ToO observation and an anonymous referee for his/her useful comments. 
%\end{acknowledgments}


%\vspace{5mm}
\facilities{\xmm, INTEGRAL}


\software{SAS
(v19.1.0; \citealt{gabriel04}), HEASoft package (v.6.31;
\citealt{ftools14}), Xspec
(v12.13.0c; \citealt{arnaud96}), xscat (v1.0.0; \citealt{smith16}), dustmaps (v1.0.4; \citealt{green18}). }



%\end{document}

%\vspace{2cm}

\appendix
\section{X-ray sources in the \grb\ field}
\label{images}

We reduced and filtered the EPIC data using standard procedures as described in Section \ref{sec:xmm} and  created pn+MOS images in three energy bands: 0.5--2 keV, 2--12 keV,  and 0.5--12 keV. These images were used to perform  source detection using the SAS tool \emph{edetect\_chain}. The resulting list of   sources was inspected visually to remove possible spurious detections and cross-correlated with optical, infrared and radio catalogs to search for possible identifications. 

\begin{figure*}
    \centering
    \includegraphics[width=\textwidth]{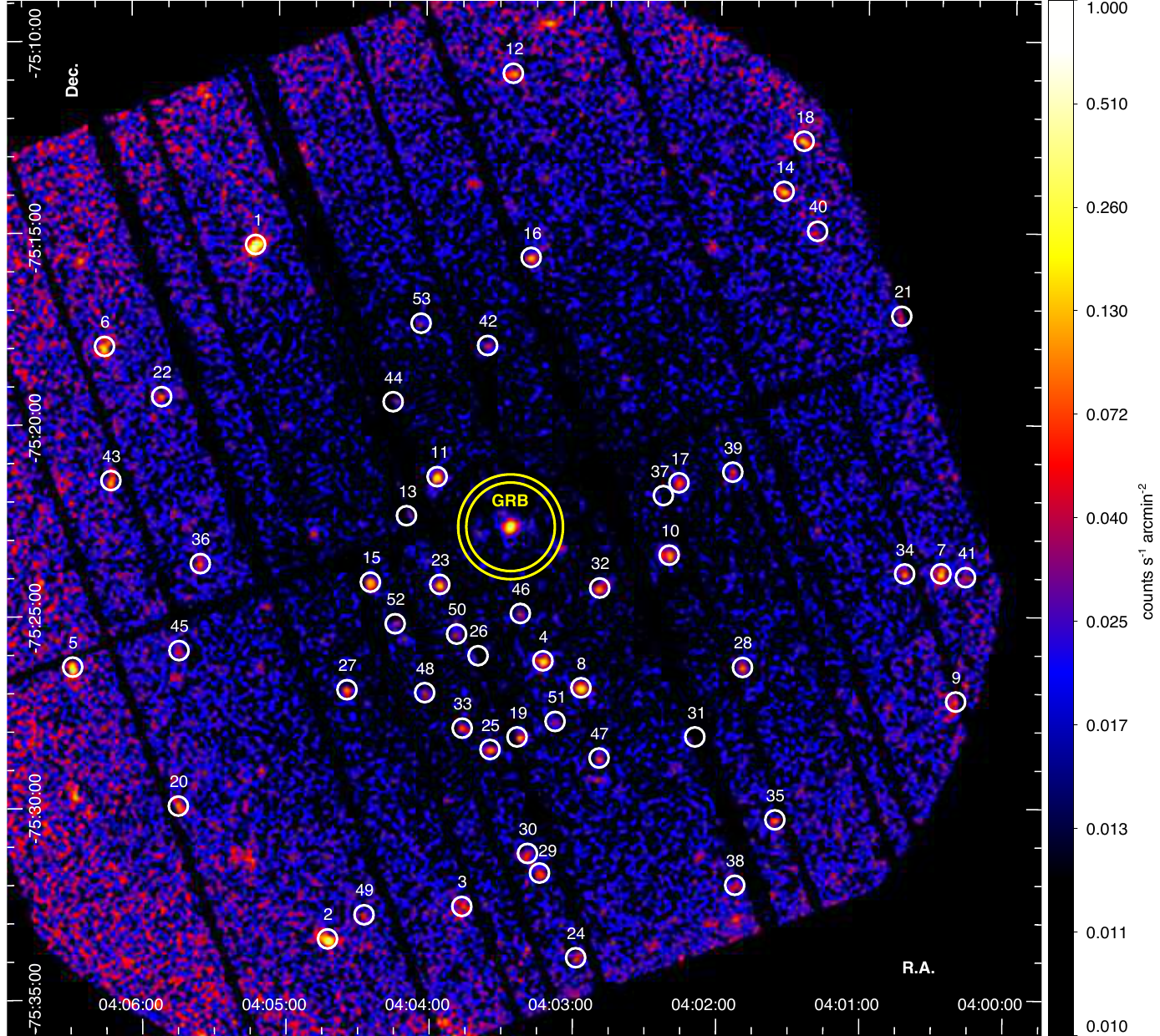}
    \caption{EPIC-pn image in the 0.3--10 keV energy range of the field of \grb .  The yellow circles show the expected position of scattering rings produced by dust in the Magellanic Bridge, for distances of 50 kpc (outermost circle) and 70 kpc (innermost circle). The sources listed in Table \ref{tab:data} are indicated by the small circles.}
    \label{fig:image}
\end{figure*}

The properties of the most interesting sources (i.e. the brightest ones and those with a plausible identification) are reported in Table \ref{tab:data}  and are shown in Figure \ref{fig:image}.
The EPIC-pn count rates 
and fluxes, derived assuming a power-law model with  photon index $\Gamma=1.7$ and a column density   $N_{\rm H}^{\rm LOC}=9.4\times10^{20}$~cm$^{-2}$  \citep{2016A&A...594A.116H}, refer to the 0.5--2 keV  range. The X-ray to optical flux ratio has been computed as

\begin{equation}
    \log (f_X/f_{\rm opt})= \log f_X+\frac{m}{2.5}+5.37,
\end{equation}

\noindent 
where  $f_X$ is the flux in the 0.5--12 keV range and  $m$ is magnitude in the  red band. When no optical counterpart was present in the error region, a limiting magnitude $m=21$ was assumed. 

Eight sources are identified with foreground stars on the basis of their $f_X/f_{\rm opt}$ and positional coincidence with objects of the GAIA and USNO-B1.0 catalogs. Several sources have been classified as quasars based on  cross-correlation with QSO candidates catalogs  \citep{2018A&A...618A.144G,2019RAA....19...29L}.

 \begin{longrotatetable}
\centerwidetable
\begin{deluxetable*}{c c  c  c  c  c  c  c  c  c}
\tabletypesize{\footnotesize}
\tablecolumns{10}
\tablecaption{The brightest or classified detected sources with X-ray and optical parameters}
\tablehead{
  \colhead{Source} &    \colhead{R.A.} & \colhead{Dec.} & \colhead{Error} & \colhead{Rate} & \colhead{Flux ($10^{-14}$} & \colhead{Class} & \colhead{Rmag} & \colhead{X/O ratio} & \colhead{Counterpart Name} \\
  \colhead{number} &  \colhead{} & \colhead{} & \colhead{(arcsec)} & \colhead{($10^{-3}$ cts s$^{-1}$)} & \colhead{erg cm$^{-2}$ s$^{-1}$)} & \colhead{}  & \colhead{mag} & \colhead{(log10)} & \colhead{}
}
\startdata
1 & 04:05:10.43 & -75:15:21.14 & 0.3 & $25.80\pm1.38$ & $3.54^{+0.32^{*}}_{-0.35}$ & QSO &  18.8 & $0.14\pm0.02$ & USNO-B1.0 0147-0023343  \\
2 & 04:04:42.37 & -75:33:28.00 & 0.4 & $14.31\pm1.13$ & $2.18^{+0.28^{*}}_{-0.33}$ & Unc. &  20.5 & $0.61\pm0.03$ & USNO-B1.0 0144-0021053  \\
3 & 04:03:46.24 & -75:32:37.90 & 2.2 & $9.59\pm1.32$ & $1.99\pm0.27$ & Unc. &  ... & $>0.59$ & ... ...  \\
4 & 04:03:12.50 & -75:26:13.62 & 0.4 & $9.41\pm0.63$ & $1.96\pm0.13$ & star &  14.2 & $-2.33\pm0.04$ & Gaia DR3 4628749515476145792  \\
5 & 04:06:27.49 & -75:26:19.46 & 0.7 & $11.24\pm1.14$ & $1.78^{+0.27^{*}}_{-0.31}$ & Unc. &  19.6 & $0.21\pm0.04$ & USNO-B1.0 0145-0021710  \\
6 & 04:06:12.80 & -75:17:58.48 & 0.6 & $7.19\pm0.88$ & $1.49\pm0.18$ & QSO &  18.9 & $-0.39\pm0.06$ & USNO-B1.0 0147-0023428  \\
7 & 04:00:27.96 & -75:23:53.16 & 0.6 & $6.82\pm0.74$ & $1.42\pm0.15$ & Unc. &  19.3 & $-0.32\pm0.06$ & USNO-B1.0 0146-0022065  \\
8 & 04:02:56.71 & -75:26:55.67 & 0.4 & $8.53\pm0.61$ & $1.25^{+0.14^{*}}_{-0.16}$ & Unc. &  20.7 & $0.31\pm0.04$ & USNO-B1.0 0145-0021402  \\
9 & 04:00:21.17 & -75:27:13.32 & 0.9 & $5.77\pm0.92$ & $1.20\pm0.19$ & star &  15.7 & $-1.88\pm0.09$ & Gaia DR3 4628795798043612800  \\
10 & 04:02:20.38 & -75:23:27.74 & 0.5 & $5.69\pm0.50$ & $1.18\pm0.10$ & Unc. &  ... & $>0.32$ & ... ...  \\
11 & 04:03:56.23 & -75:21:25.54 & 0.4 & $8.10\pm0.59$ & $1.15^{+0.15^{*}}_{-0.17}$ & QSO &  19.5 & $-0.18\pm0.03$ & USNO-B1.0 0146-0022388  \\
12 & 04:03:24.89 & -75:10:54.67 & 0.9 & $5.09\pm0.72$ & $1.06\pm0.15$ & Unc. &  19.2 & $-0.45\pm0.07$ & USNO-B1.0 0148-0024169  \\
13 & 04:04:08.92 & -75:22:26.51 & 0.6 & $4.85\pm1.19$ & $1.01\pm0.25$ & Unc. &  ... & $>0.22$ & ... ...  \\
14 & 04:01:34.12 & -75:13:57.52 & 0.6 & $4.80\pm0.65$ & $1.00\pm0.13$ & Unc. &  20.1 & $-0.06\pm0.06$ & Gaia DR3 4628825240041806080  \\
15 & 04:04:23.92 & -75:24:10.33 & 0.5 & $6.52\pm0.56$ & $0.98^{+0.15^{*}}_{-0.17}$ & QSO &  19.9 & $-0.10\pm0.04$ & Gaia DR3 4628772532202907008  \\
16 & 04:03:17.55 & -75:15:42.63 & 0.7 & $4.33\pm0.54$ & $0.90\pm0.11$ & Unc. &  20.5 & $-0.05\pm0.06$ & Gaia DR3 4628821460470380928  \\
17 & 04:02:16.49 & -75:21:34.49 & 0.5 & $4.23\pm0.45$ & $0.88\pm0.09$ & Unc. &  ... & $>0.10$ & ... ...  \\
18 & 04:01:26.12 & -75:12:38.80 & 0.5 & $5.87\pm0.78$ & $0.88^{+0.19^{*}}_{-0.24}$ & QSO &  19.2 & $-0.27\pm0.06$ & Gaia DR3 4628825313058849024  \\
19 & 04:03:23.21 & -75:28:12.74 & 0.6 & $6.25\pm0.80$ & $0.84^{+0.14^{*}}_{-0.17}$ & Unc. &  19.7 & $-0.19\pm0.06$ & USNO-B1.0 0145-0021446  \\
20 & 04:05:44.22 & -75:29:58.64 & 0.9 & $3.98\pm0.72$ & $0.83\pm0.15$ & QSO &  18.2 & $-0.91\pm0.08$ & Gaia DR3 4628768099796605056  \\
21 & 04:00:45.39 & -75:17:11.12 & 0.9 & $3.66\pm0.68$ & $0.76\pm0.14$ & Unc. &  ... & $>0.11$ & ... ...  \\
22 & 04:05:49.51 & -75:19:17.97 & 1.1 & $3.58\pm0.60$ & $0.74\pm0.12$ & star &  11.0 & $-3.94\pm0.09$ & Gaia DR3 4628771776288001536  \\
23 & 04:03:55.22 & -75:24:14.16 & 0.5 & $5.45\pm0.50$ & $0.73^{+0.12^{*}}_{-0.14}$ & Unc. &  20.5 & $0.01\pm0.04$ & USNO-B1.0 0145-0021502  \\
24 & 04:02:58.65 & -75:33:57.98 & 1.6 & $3.41\pm0.64$ & $0.71\pm0.13$ & star &  10.4 & $-4.56\pm0.19$ & Gaia DR3 4628746251297682816  \\
25 & 04:03:34.46 & -75:28:32.39 & 0.6 & $3.35\pm0.43$ & $0.70\pm0.09$ & Unc. &  ... & $>0.08$ & ... ...  \\
26 & 04:03:39.39 & -75:26:05.71 & 0.7 & $3.24\pm0.80$ & $0.67\pm0.17$ & Unc. &  ... & $>-0.32$ & ... ...  \\
27 & 04:04:33.81 & -75:26:58.40 & 0.8 & $3.23\pm0.43$ & $0.67\pm0.09$ & Unc. &  ... & $>0.07$ & ... ...  \\
28 & 04:01:49.74 & -75:26:22.71 & 0.8 & $3.20\pm0.43$ & $0.66\pm0.09$ & QSO &  20.2 & $-0.23\pm0.06$ & Gaia DR3 4628795179565576832  \\
29 & 04:03:13.81 & -75:31:45.63 & 1.1 & $3.19\pm0.51$ & $0.66\pm0.11$ & Unc. &  ... & $>0.04$ & ... ...  \\
30 & 04:03:18.83 & -75:31:14.98 & 1.1 & $3.14\pm0.58$ & $0.65\pm0.12$ & Unc. &  ... & $>0.17$ & ... ...  \\
31 & 04:02:09.33 & -75:28:11.85 & 1.2 & $3.06\pm1.20$ & $0.64\pm0.25$ & Unc. &  ... & $>0.13$ & ... ...  \\
32 & 04:02:49.05 & -75:24:19.52 & 0.5 & $4.38\pm0.43$ & $0.63^{+0.10^{*}}_{-0.12}$ & Unc. &  ... & $>0.14$ & ... ...  \\
33 & 04:03:46.01 & -75:27:59.01 & 0.8 & $2.98\pm0.46$ & $0.62\pm0.10$ & star &  15.9 & $-2.10\pm0.08$ & USNO-B1.0 0145-0021480  \\
34 & 04:00:42.97 & -75:23:54.06 & 1.0 & $2.93\pm0.52$ & $0.61\pm0.11$ & Unc. &  ... & $>-0.01$ & ... ...  \\
35 & 04:01:35.81 & -75:30:20.66 & 0.8 & $2.67\pm0.50$ & $0.55\pm0.10$ & Unc. &  ... & $>0.06$ & ... ...  \\
36 & 04:05:34.19 & -75:23:39.68 & 1.0 & $2.55\pm0.49$ & $0.53\pm0.10$ & Unc. &  ... & $>0.02$ & ... ...  \\
37 & 04:02:22.90 & -75:21:54.58 & 0.8 & $2.35\pm0.69$ & $0.49\pm0.14$ & QSO &  18.0 & $-1.49\pm0.24$ & USNO-B1.0 0146-0022230  \\
38 & 04:01:52.33 & -75:32:03.49 & 1.6 & $2.26\pm0.48$ & $0.47\pm0.10$ & Unc. &  ... & $>-0.15$ & ... ...  \\
39 & 04:01:54.32 & -75:21:17.31 & 0.8 & $2.16\pm0.36$ & $0.45\pm0.07$ & Unc. &  ... & $>-0.12$ & ... ...  \\
40 & 04:01:20.27 & -75:14:59.55 & 1.3 & $2.15\pm0.48$ & $0.45\pm0.10$ & Unc. &  19.8 & $-0.58\pm0.12$ & USNO-B1.0 0147-0022964  \\
41 & 04:00:17.79 & -75:23:58.63 & 1.9 & $2.11\pm0.50$ & $0.44\pm0.10$ & star &  12.3 & $-3.67\pm0.13$ & Gaia DR3 4628796038561772928  \\
42 & 04:03:35.41 & -75:18:00.50 & 0.8 & $2.10\pm0.35$ & $0.44\pm0.07$ & Unc. &  ... & $>-0.32$ & ... ...  \\
43 & 04:06:10.78 & -75:21:28.39 & 0.9 & $2.08\pm0.63$ & $0.43\pm0.13$ & QSO &  20.1 & $-0.41\pm0.13$ & USNO-B1.0 0146-0022585  \\
44 & 04:04:14.29 & -75:19:28.19 & 1.0 & $2.08\pm0.59$ & $0.43\pm0.12$ & Unc. &  ... & $>-0.25$ & ... ...  \\
45 & 04:05:43.35 & -75:25:55.43 & 1.5 & $1.92\pm0.47$ & $0.40\pm0.10$ & Unc. &  ... & $>-0.23$ & ... ...  \\
46 & 04:03:21.87 & -75:24:59.60 & 0.9 & $1.92\pm0.32$ & $0.40\pm0.07$ & Unc. &  ... & $>-0.31$ & ... ...  \\
47 & 04:02:49.10 & -75:28:45.49 & 1.1 & $1.88\pm0.34$ & $0.39\pm0.07$ & Unc. &  ... & $>-0.19$ & ... ...  \\
48 & 04:04:01.56 & -75:27:03.37 & 0.8 & $1.75\pm0.33$ & $0.36\pm0.07$ & Unc. &  ... & $>-0.29$ & ... ...  \\
49 & 04:04:27.08 & -75:32:50.58 & 1.0 & $1.74\pm0.46$ & $0.36\pm0.10$ & QSO &  20.0 & $-0.63\pm0.15$ & Gaia DR3 4628745014347539968  \\
50 & 04:03:48.28 & -75:25:31.90 & 0.9 & $1.62\pm0.34$ & $0.34\pm0.07$ & Unc. &  ... & $>-0.28$ & ... ...  \\
51 & 04:03:07.43 & -75:27:48.46 & 1.0 & $1.43\pm0.31$ & $0.30\pm0.06$ & Unc. &  ... & $>-0.50$ & ... ...  \\
52 & 04:04:13.72 & -75:25:15.45 & 1.4 & $1.23\pm0.29$ & $0.26\pm0.06$ & star &  18.5 & $-1.46\pm0.12$ & USNO-B1.0 0145-0021522  \\
53 & 04:04:02.71 & -75:17:25.46 & 1.2 & $1.05\pm0.29$ & $0.22\pm0.06$ & star &  19.5 & $-1.05\pm0.13$ & USNO-B1.0 0147-0023232  \\
\enddata
\label{tab:data}
\end{deluxetable*}
\end{longrotatetable}

 % \input{table_sources_2}
 %   \input{table_sources}

%\section{Appendix information}

%Appendices can be broken into separate sections just like in the main text.
%the section letter appended.  Here is an equation as an example.
%\begin{equation}
%I = \frac{1}{1 + d_{1}^{P (1 + d_{2} )}}
%\end{equation}
%Appendix tables and figures should not be numbered like equations. Instead
  

%\bibliography{grb}

\bibliographystyle{aasjournal}



\end{document}